\documentclass[sigconf]{acmart}

\AtBeginDocument{%
  \providecommand\BibTeX{{%
    \normalfont B\kern-0.5em{\scshape i\kern-0.25em b}\kern-0.8em\TeX}}}

\copyrightyear{2021}
\acmYear{2021}
\setcopyright{acmlicensed}\acmConference[CIKM '21]{Proceedings of the 30th ACM International Conference on Information and Knowledge Management}{November 1--5, 2021}{Virtual Event, QLD, Australia}
\acmBooktitle{Proceedings of the 30th ACM International Conference on Information and Knowledge Management (CIKM '21), November 1--5, 2021, Virtual Event, QLD, Australia}
\acmPrice{15.00}
\acmDOI{10.1145/3459637.3482093}
\acmISBN{978-1-4503-8446-9/21/11}

\usepackage{multirow}
\usepackage{graphicx}
\newcommand{\reviewer}[3]{
	\expandafter\newcommand\csname #1\endcsname[1]{
		\textcolor{#3}{[#2: ##1]}
	}
}
\reviewer{yg}{YG}{blue}

\begin{document}

\fancyhead{}
\title{Dual Correction Strategy for Ranking Distillation \\in Top-N Recommender System}

\author{Youngjune Lee,\textsuperscript{\rm 1} Kee-Eung Kim\textsuperscript{\rm 1, 2}}

\affiliation{%
  \institution{\textsuperscript{\rm 1}{School of Computing}, KAIST, Daejeon \country{Republic of Korea}}
  \institution{\textsuperscript{\rm 2}Graduate School of AI, KAIST, Daejeon \country{Republic of Korea}}
}

\email{{dudwns511, kekim}@kaist.ac.kr}
\begin{abstract}
Knowledge Distillation (KD), which transfers the knowledge of a well-trained large model (teacher) to a small model (student), has become an important area of research for practical deployment of recommender systems. 
Recently, Relaxed Ranking Distillation (RRD) has shown that distilling the ranking information in the recommendation list significantly improves the performance.
However, the method still has limitations in that 
\textbf{1)} it does not fully utilize the prediction errors of the student model, which makes the training not fully efficient, and \textbf{2)} it only distills the user-side ranking information, which provides an insufficient view under the sparse implicit feedback.
This paper presents Dual Correction strategy for Distillation (DCD), which transfers the ranking information from the teacher model to the student model in a more efficient manner.
Most importantly, DCD uses the \textit{discrepancy} between the teacher model and the student model predictions to
decide which knowledge to be distilled.
By doing so, DCD essentially provides the learning guidance tailored to ``correcting'' 
what the student model has failed to accurately predict.
This process is applied for transferring the ranking information from the user-side as well as
the item-side to address sparse implicit user feedback. 
Our experiments show that the proposed method outperforms the state-of-the-art baselines, and ablation studies validate the effectiveness of each component.
\end{abstract}



\maketitle

\section{Introduction}
In this era of information explosion, Recommender Systems (RS) are widely used in various industries to provide personalized user experience \cite{ProxySR, SSCDR, BUIR, DRE}.
For achieving higher recommendation accuracy, the recommendation model has become very large
to capture the complexity of personalized recommendations~\cite{CD,RD, BD, GCN_distill,DCF, HTD}.
However, large models incur correspondingly large computational cost as well as high latency for inference,
which has become one of the major obstacles for real-time service~\cite{DERRD, BD, HTD}.

To reduce the inference latency, early methods adopt hash techniques \cite{candidategeneration,hash2,hash1,DCF} or tree-based data structure  \cite{tree_RS}.
However, they have problems such as easily falling into a local optimum or applicable only to specific models \cite{DERRD, GCN_distill}.
To address the problems, Knowledge Distillation (KD) has been actively studied for RS \cite{DERRD, RD, CD, BD, GCN_distill, IRRRD}.
KD is a model compression technique that improves the performance of a small student model by transferring the knowledge of a pre-trained large teacher model \cite{chen2017learning,self_distill1,KD,FitNet}.
During the distillation, the teacher model provides additional supervision which is not 
existent in the users' feedback, so the student model can achieve a higher recommendation accuracy compared to the student model trained only on the original feedback.

The state-of-the-art method, Relaxed Ranking Distillation (RRD) \cite{DERRD}, formulates the distillation process as a \textit{ranking matching problem} between the recommendation list of the teacher model and that of the student model. In other words, it utilizes the ranking orders among the items from the teacher model as additional supervision to guide the student model, and trains the student model to preserve the ranking orders of the teacher model.
This ranking-distillation approach transfers the relative preference order among the user's preferred items, which is the key knowledge directly affecting top-$N$ recommendation accuracy.
As a result, it significantly improves performance over the previous methods \cite{RD, CD} that do not directly utilize the ranking information \cite{DERRD}.

Still, there are limitations in RRD.
First, it transfers the knowledge without consideration of the prediction errors of the student model.
As the student model gets more accurate in matching the prediction of the teacher model, repeatedly distilling the ranking information that the student model already correctly predicts cannot effectively enhance the student model and makes the training inefficient. 
We argue that the knowledge to be distilled should be dynamically changed based on the student model’s prediction error, enabling “correction” for what the student model has not yet predicted accurately.
Second, it only transfers user-side ranking information, i.e., ranking orders among the items.
Previous studies have pointed out that learning only with the user-side ranking information degrades the quality of user representation \cite{multi-object} and provides a view insufficient to fully understand the sparse implicit feedback \cite{dual_NPR, IRRRD}.
Particularly in KD where the student model’s capacity is limited, 
these problems can be further exacerbated and severely degrade the performance.

In this work, we propose a novel Dual Correction strategy for Distillation (DCD), 
which aims to address the aforementioned shortcomings.
To this end, DCD first computes \textit{discrepancy} between the ranking list of the teacher model and that of the student model, then decides what knowledge to be distilled based on the discrepancy.
By doing so, DCD provides guidance tailored to correct what the student model has failed to correctly predict, which helps to find an effective path for the student model’s training.
This process is conducted for dual-side ranking, i.e., for the user-side and the item-side, providing a comprehensive view to better understand both users and items \cite{dual_NPR, multi-object}.
We validate the superiority of the proposed method with extensive experiments on real-world datasets,
and provide an ablation study showing the effectiveness of each proposed component.

\section{Methodology}
\subsection{Problem formulation and Notations}
We focus on top-$N$ recommendation task for implicit feedback \cite{hu2008collaborative, BUIR}.
Given implicit user-item interactions, a recommender system provides a ranked list of top-$N$ unobserved items for each user. 
The distillation process is conducted as follows:
First, we train a large model (teacher) using the implicit feedback.
Then, we train a small model (student) with the same feedback data along with the ranking list predicted from the teacher. 
The ranking information reveals the detailed preference orders among the unobserved items, which helps the training of the student.
Our goal is to design a distillation strategy that allows the student to effectively follow the teacher's ranking list.

We denote the teacher by $T$ and the student by $S$.
$R_T^u$ and $R_S^u$ denote the user-side ranking list for user $u$ (i.e., the list of the unobserved items) predicted by the teacher and the student, respectively. 
$R_*^{u}(i)$ denotes the rank of item $i$ in the ranking list where a lower value means a higher ranking position, i.e., $R_*^{u}(i)=0$ is the highest ranking.
For the item-side ranking list, we simply reverse the notation of the user-side.
Concretely, $R_T^i$ and $R_S^i$ denote the item-side ranking list for item $i$ (i.e., the list of the unobserved users) predicted by the teacher and the student, respectively, and $R_*^{i}(u)$ denotes the rank of user $u$ in the ranking list.

\subsection{Distilling the Ranking Information}
The ranking distillation (RRD) \cite{DERRD} formulates the distillation as a ranking matching problem between the ranking list of the teacher and that of the student (i.e., $R_T^u$ and $R_S^u$).
Specifically, the method trains the student to preserve the orders of ranking in $R_T^u$ by using a variant of ListMLE \cite{xia2008list-wise}.
The core idea is to define a permutation 
probability based on the the student's ranking scores, and train the student to maximize the likelihood of the teacher's ranking $R_T^u$.

To make the student better focus on top-ranked items, we also adopt \textit{relaxed permutation probability} \cite{DERRD} that ignores the low-ranked items' detailed orders.
Formally, let $R_T^u$ is decomposed to two sub-ranking lists $R_T^u = [\pi; \pi']$, where $\pi$ includes a few top-ranked items and $\pi'$ includes the remaining items ($u$ and $T$ are omitted for simplicity).
The relaxed permutation probability of the ranked list $\pi$ for the student $S$ is defined as follows:
 \begin{equation}
 \label{RRD_prob}
     p(\pi|S) =  \prod_{{k}={1}}^{|\pi|} 
     \frac{{\exp} \, S\left(u, \pi_k \right) }
     {\sum_{{i}= {k}}^{|\pi|}  {\exp} \, S\left(u, \pi_i \right)
     + {\sum_{{j}= {1}}^{|\pi'|} {\exp} \, S(u, \pi'_j )}},
 \end{equation}
where $\pi_k$ is the $k$-th item in $\pi$, $S(u, \pi_k)$ is the score of the user-item interaction predicted by the student. 
By maximizing the probability, the student learns the detailed ranking orders in $\pi$ while 
lowering all the ranks of items in $\pi'$ below the lowest rank of items in $\pi$, which allows the student to focus more on top-$N$ ranking orders. 
The student is trained by the ranking knowledge distillation (RKD) loss as follows:
\begin{equation}
    \min \mathcal{L}_{RKD} = \mathcal{L}_{RS} + \lambda_{RRD} \mathcal{L}_{RRD},
\end{equation}
where $\mathcal{L}_{RS}$ is the loss for training the base model using the implicit feedback data,
and 
$\mathcal{L}_{RRD}=-{1\over|B|}\sum_{u \in {B}} \log {p}(\pi^u | S)$ is the permutation loss 
defined for the users in mini-batch $B$.

\subsection{Dual Correction Strategy}
We present Dual Correction strategy for Distillation (DCD) that adaptively assigns more 
concentrations on training instances that the student fails to predict correctly, unlike the prior methods such as
RRD that generate training instances solely based on the teacher's predictions.
This correction is used for transferring the ranking information from the user-side as well as the item-side, providing a comprehensive view to understand both users and items.

\subsubsection{\textbf{Identifying discrepancy between Teacher model and Student model}}
\label{subsection:student_aware}
DCD first identifies \textit{discrepancy} between $R^u_T$ and $R^u_S$ to decides what knowledge to be distilled.
We define two types of discrepancy: 1) underestimation error and 2) overestimation error.
The underestimation error means that the student predicts a low ranking position whereas the teacher predicts a higher ranking position, i.e., $R^u_S(i) > R^u_T(i)$.
Thus, the student needs to be corrected to give a lower rank value for $(u,i)$.
The overestimation error means the opposite, the student predicts a high ranking position whereas the teacher predicts a lower ranking position, i.e., $R^u_S(i) < R^u_T(i)$, which needs to be corrected to give a higher rank value.
The user-side errors are computed as follows:
\begin{equation}
\begin{aligned}
\label{rank_discrepancy}
   D^{u}_{i}(S,T) =   {tanh}({max}({\mu}({R}_{S}^{u}(  {i})-{R}_{T}^{u}({i})),\mathbf{0}))
\end{aligned}
\end{equation}
\begin{equation}
\begin{aligned}
\label{rank_discrepancy}
   D^{u}_{i}(T,S) =   {tanh}({max}({\mu}({R}_{T}^{u}(  {i})-{R}_{S}^{u}({i})),\mathbf{0})) 
\end{aligned}
\end{equation}
We use $tanh$, which is a saturated function, to treat the errors above a certain threshold equally, allowing the student to learn the teacher’s knowledge on most of the discrepant predictions.
$\mu$ is a hyperparameter that controls the sharpness of the tanh function.
Using the computed errors, we identify the discrepant predictions that need to be corrected.
Concretely, we sample $M^u_l$ underestimated items and $M^u_h$ overestimated items.
Both sampling probabilities are proportional to the degree of a discrepancy.
\begin{equation}
\label{user_side_rank_difference}
{p}^{u}_{l}(i)\propto  D^{u}_{i}(S,T), \quad {p}^{u}_{h}(i)\propto  D^{u}_{i}(T,S)
\end{equation}
where ${p}^{u}_{l}(i)$ is the sampling probability of underestimated items and ${p}^{u}_{h}(i)$ is the sampling probability of overestimated items for user $u$.
These discrepant items are dynamically changed based on the prediction errors of the student during the training, and will be corrected by the correction loss (Sec. \ref{subsection:DCD}).

DCD also provides the corrections for discrepancy in terms of the item-side ranking. 
As pointed out in the previous work \cite{multi-object, dual_NPR}, learning only the user-side ranking degrades the quality of user representation \cite{multi-object} and also provides a restricted view insufficient to understand the sparse implicit feedback \cite{dual_NPR}. 
Especially, in KD where the student’s capacity is highly limited, these problems can be further exacerbated, which leads to degraded performance.

Similar to the user-side, we identify the discrepant predictions on the item-side.
We sample $M^i_l$ underestimated users and $M^i_h$ overestimated users based on the discrepancy between $R^i_T$ and $R^i_S$.
The sampling probabilities are as follows:
\begin{equation}
\label{item_side_prob}
{p}^{i}_{l}(u)\propto  D^{i}_{u}(S,T), \quad {p}^{i}_{h}(u)\propto  D^{i}_{u}(T,S),
\end{equation}
where ${p}^{i}_{l}(u)$ is the probability of underestimated users and ${p}^{i}_{h}(u)$ is the probability of overestimated users for item $i$.
Without loss of generality, $D^{i}_{u}(A,B) =  {tanh}({max}({\mu}({R}_{A}^{i}({u})-{R}_{B}^{i}({u})),\mathbf{0}))$.

\subsubsection{\textbf{Dual Correction Distillation Loss}}
\label{subsection:DCD}
Now, we correct the  discrepant predictions in the user-side (summarized by $M^u_l$ and $M^u_h$) and the item-side (summarized by $M^i_l$ and $M^i_h$).
From the points of the teacher, the underestimation errors contain the predictions that should be higher-ranked, whereas the overestimation errors contain the predictions that should be relatively lower-ranked compared to the former.
As consistently shown in the existing distillation work \cite{DERRD, RD, CD}, the student takes a huge benefit by learning the teacher's knowledge with a particular emphasis on the high-ranked items, because it directly affects the top-$N$ recommendation accuracy.  
In this regard, we design the correction loss that corrects the ranks of the underestimation errors in detail and lowers the ranks of the overestimation errors overall.

Let $\rho^u$ denote the sorted lists of $M^u_l$ by the original order in $R^u_T$, and $\rho^i$ denote the sorted lists of $M^i_l$ by the order in $R^i_T$.
The user-side correction distillation (UCD) for user $u$ is conducted by maximizing the following relaxed permutation probability:
 \begin{equation}
 \label{UCD_prob}
     p(\rho^u|S) =  \prod_{{k}={1}}^{|\rho^u|} 
     \frac{{\exp} \, S(u, \rho^u_k ) }
     {\sum_{{i}= {k}}^{|\rho^u|}  {\exp} \, S(u, \rho^u_i )
     + {\sum_{j \in M^u_h} {\exp} \, S(u, j )}},
 \end{equation}
where $\rho^u_k$ is the $k$-th item in $\rho^u$.
UCD is applied for the users in mini-batch $B$, i.e., $\mathcal{L}_{UCD}=-{1\over|B|}\sum_{u \in {B}} \log {p}(\rho^u| S)$. 
Analogously, item-side correction distillation (ICD) for item $i$ is conducted by maximizing the following relaxed permutation probability:
 \begin{equation}
 \label{UCD_prob}
     p(\rho^i|S) =  \prod_{{k}={1}}^{|\rho^i|} 
     \frac{{\exp} \, S(\rho^i_k, i) }
     {\sum_{{j}= {k}}^{|\rho^i|}  {\exp} \, S(\rho^i_j, i)
     + {\sum_{l \in M^i_h} {\exp} \, S(l, i)}}.
 \end{equation}
ICD is also applied for correcting errors with respect to the items in the batch, i.e., $\mathcal{L}_{ICD}=-{1\over|B|}\sum_{i \in {B}} \log {p}(\rho^i| S)$. 
Finally, the proposed DCD trains the student with the following loss function.
\begin{equation}\label{DCR}
 \min_{\theta_s} \mathcal{L}_{RKD} +\lambda_{UCD} \mathcal{L}_{UCD} + \lambda_{ICD}\mathcal{L}_{ICD},
\end{equation}
where $\theta_s$ is the learning parameters of the student.
$\lambda_{UCD}$ and $\lambda_{ICD}$ are hyperparameters controlling the user-side and item-side corrections, respectively. 
Note that $\mathcal{L}_{RKD}$ is computed for the same ground-truths regardless of the discrepancy during the training.
Finally, our dual correction loss provides dynamically changing guidance to correct the student errors for more effective training.

\begin{table}[h]
\caption{The number of parameters and inference time for generating recommendation list for every user.}
\vspace{-10pt}
\centering
\renewcommand{\tabcolsep}{1.2mm}
\renewcommand{\arraystretch}{0.5}
\begin{tabular}{cccc}
\toprule
Dataset                     & Base Model          & \# Parameters & Time (GPU/CPU)    \\ \midrule
\multirow{5}{*}{CiteULike}  & BPR (Teacher)      & 6.08M      & 5.90s / 59.80s     \\  
                            & BPR (Student)            & 0.61M      & 3.67s / 14.90s    \\ \cmidrule{2-4} 
                            & NeuMF (Teacher)   & 12.24M     & 17.80s / 261.37s  \\ 
                            & NeuMF (Student)             & 1.22M      & 7.42s / 33.75s    \\ \midrule
\multirow{5}{*}{Foursquare} & BPR (Teacher)      & 9.61M      & 25.22s / 252.23s  \\  
                            &  BPR (Student)           & 0.96M      & 15.22s / 64.22s   \\ \cmidrule{2-4} 
                            & NeuMF (Teacher)     & 19.30M     & 75.22s / 1086.44s \\  
                            &  NeuMF (Student)         & 1.92M      & 30.54s / 144.05s   \\ \toprule
\end{tabular}
\label{table:time}
\vspace*{-0.7cm}
\end{table}
\section{Experiments}

\subsection{Experiment Setup}

We closely follow the setup of the state-of-the-art method, RRD \cite{DERRD}.
Specifically, datasets, base models, evaluation protocol, and the metrics are the same as \cite{DERRD}.
Due to the limited space, we omit the detailed explanations of the setup.
Please refer to \cite{DERRD}.

\noindent
\textbf{Datasets.} We use CiteULike \cite{CiteULike} and Foursquare \cite{Foursquare} which are public real-world datasets.
After the preprocessing \cite{DERRD}, CiteULike has 5,220 users, 25,182 items, and 115,142 interactions. 
Foursquare has 19,466 users, 28,594 items, and 609,655 interactions.

\noindent
\textbf{Base models.} 
We use two base models for the top-$N$ recommendation: BPR \cite{BPR} and NeuMF \cite{NeuMF}, which have different architectures and optimization strategies.
For both models, the dimension of user/item representations are set to 200 for the teacher model, and 20 for the student model.
Following \cite{DERRD}, we denote the student model trained without distillation as "Student".
Table \ref{table:time} presents the number of parameters and inference time. 
The inferences are made using PyTorch with CUDA from TITAN Xp GPU and Intel i7-4770 CPU. 
It shows that the smaller model has lower inference latency.

\noindent
\textbf{Evaluation protocol and metrics.}
We use the leave-one-out evaluation protocol whereby two interacted items for each user are held out for test/validation, and the rest are used for training \cite{DERRD}.
We adopt two top-$N$ ranking evaluation metrics, namely Hit Ratio (H@$N$) and Mean Reciprocal Rank (M@$N$).
We compute the average score of those two metrics for each user. Finally, we report the average of the five independent runs.

\noindent
\textbf{Baselines.}
We compare DCD with the state-of-the-art ranking distillation method, RRD \cite{DERRD}.
Note that we do not include the previous methods distilling point-wise information (e.g., RD \cite{RD}, CD \cite{CD}), because RRD already outperforms them by a huge margin \cite{DERRD}.  

\noindent
\textbf{Implementation details.}
We use PyTorch \cite{paszke2019pytorch} for implementation and train all models with Adam optimizer \cite{kingma2014adam}.
For each base model and dataset, we tune the hyperparameters by grid search on the validation set.
We tune learning rate and L2 regularizer $\in \{10^{-1}, 10^{-2}, 10^{-3},$ $ 10^{-4}, 10^{-5}, 10^{-6}\}$.
In the case of RRD-specific hyperparameters, we tune them in the ranges suggested by the original paper.
For the dual correction loss, 
we tune $\lambda_{UCD}, \lambda_{ICD} \in \{1, 10^{-1}, 10^{-2}, 10^{-3}, 10^{-4}, 10^{-5}\}$, 
and conduct the sampling process every 5 epochs.
The number of discrepant users/items ($M^*_l$,$M^*_h$) is set to 40, but it can be further tuned.
Lastly, $\mu$ is set to $10^{-3}$.


\begin{table}[t]
\caption{Performance comparison. \textit{improve.r} denotes the improvement of DCD over RRD and \textit{improve.s} denotes the
improvement of DCD over Student. * and ** indicate $p$ $\le$ 0.005 and $p$ $\le$ 0.0005 for the paired t-test of DCD vs. RRD on H@5.}
\vspace{-10pt}
\centering
\renewcommand{\arraystretch}{0.2}
\renewcommand{\tabcolsep}{1.2mm}
\textbf{CiteULike}
\begin{tabular}{clcccc}
\toprule
 Base Model & \multicolumn{1}{c}{KD Method}    & H@5  & M@5 & H@10  & M@10           \\ \midrule
 \multirow{18}{*}{BPR}   & Teacher  & 0.5282  & 0.3704      & 0.6328    & 0.3844           \\  
                           & Student    & 0.4570  & 0.3061              & 0.5673          & 0.3214        \\
                           & RRD      & 0.4735    & 0.3200           & 0.5800          & 0.3343                \\ \cmidrule(l){2-6} 
                             & DCD *      & \textbf{0.4989}         & \textbf{0.3412}       & \textbf{0.6082} & \textbf{0.3560}            \\ \cmidrule(l){2-6} 
                            & \textit{improve.r }& 5.36\%          & 6.63\%               & 4.86\%          & 6.49\%         \\
                          &  \textit{improve.s} & 9.17\%          & 11.47\%           & 7.21\%          & 10.77\%     \\ \cmidrule(l){1-6}   
 \multirow{18}{*}{NeuMF} & Teacher                       & 0.4840          & 0.3346           & 0.5827          & 0.3478            \\ 
                            & Student                       & 0.3805          & 0.2499             & 0.4817          & 0.2634           \\ 
                             & RRD                           & 0.4563          & 0.2952             & 0.5647          & 0.3092        \\ \cmidrule(l){2-6}
  & DCD ** & \textbf{0.4700} & \textbf{0.3101} & \textbf{0.5742} & \textbf{0.3241}   \\ \cmidrule(l){2-6}
                            & \textit{improve.r} & 3.00\%   & 5.05\%   & 1.68\% & 4.82\%   \\
                           &  \textit{improve.s} & 23.52\%         & 24.09\%            & 19.20\%         & 23.04\%         \\ \midrule

\end{tabular}
\textbf{Foursquare}
\begin{tabular}{clcccc}
\toprule
 Base Model & \multicolumn{1}{c}{KD Method}    & H@5  & M@5   & H@10  & M@10             \\ \midrule

 \multirow{18}{*}{BPR}   & Teacher                       & 0.5623          & 0.3618           & 0.7068          & 0.3812         \\ 
                           & Student                       & 0.4982          & 0.3140            & 0.6498          & 0.3342      \\ 
                        & RRD                           & 0.5132          & 0.3259            & 0.6625          & 0.3454            \\ \cmidrule(l){2-6} 
                             & DCD **                   & \textbf{0.5468}          & \textbf{0.3483}             & \textbf{0.6958}          & \textbf{0.3683}              \\ \cmidrule(l){2-6}
                          & \textit{improve.r} & 6.55\%          & 6.87\%       & 5.03\%          & 6.63\%     \\
                    &  \textit{improve.s} & 9.76\%          & 10.92\%           & 7.08\%          & 10.20\%       \\ \cmidrule(l){1-6} 
 \multirow{18}{*}{NeuMF} & Teacher                       & 0.5459          & 0.3499             & 0.6897          & 0.3693         \\ 
                   & Student                       & 0.4793          & 0.2959                & 0.6312          & 0.3162     \\
                     & RRD                           & 0.5076          & 0.3119          & 0.6615          & 0.3313               \\ \cmidrule(l){2-6}  
                 & DCD **                  & \textbf{0.5287} & \textbf{0.3303}            & \textbf{0.6734} & \textbf{0.3497}           \\ \cmidrule(l){2-6} 
                &  \textit{improve.r} & 4.16\%          & 5.90\%         & 1.80\%          & 5.55\%           \\
                      & \textit{improve.s} & 10.31\%         & 11.63\%            & 6.69\%         & 10.59\%         \\ \bottomrule
\end{tabular}
\label{table:performance}
\vspace*{-0.3cm}
\end{table}
\vspace*{-0.1cm}

\subsection{Results}
\subsubsection{\textbf{Overall Evaluation}}
Table \ref{table:performance} presents top-$N$ recommendation accuracy of the methods compared.
DCD achieves significantly higher performance than RRD on both datasets and both base models. 
Also, in terms of the number of recommended items ($N$), DCD shows larger improvements for H@5/M@5 compared to H@10/M@10. 
Namely, DCD has a better performance at predicting the top-ranked items than RRD, which is practically advantageous for real-world RS, which gives the users the most preferred items.

\subsubsection{\textbf{Ablation Study}}
We provide ablation study of the key components of DCD in Table \ref{table:ablation}.
We compare the following ablations:
1) \textbf{w/o Correction} transfers the user-side and item-side ranking information without the correction strategy, i.e., RRD + item-side RRD.
2) \textbf{w/o Item-side} and \textbf{w/o User-side} ablate ICD and UCD from DCD, respectively.
3) \textbf{w/o Sampling} deterministically selects items/users with the largest underestimation error and overestimation error without the sampling process (Sec. \ref{subsection:student_aware}).
We observe that each proposed component is indeed effective in distilling the ranking information.
This result supports our claim that the supervision from the teacher model should be dynamically changed based on the student's errors (w/o Correction) and distilling the single-side ranking is insufficient (w/o Item-side and w/o User-side).
Also, the comparison with ``w/o Sampling'' shows that a certain degree of flexibility is beneficial in choosing the discrepant predictions for the correction strategy.

\begin{table}[t]
\renewcommand{\arraystretch}{0.6}
\renewcommand{\tabcolsep}{1.2mm}
\caption{Ablation analysis on Foursquare dataset.}
\vspace{-10pt}
\begin{tabular}{clcccc}
\toprule
Base Model             & \multicolumn{1}{c}{KD Method} & H@5 & M@5 & H@10 & M@10 \\ \midrule
\multirow{6}{*}{BPR}   & DCD     & \textbf{0.5468}         & \textbf{0.3483}         & \textbf{0.6958}          & \textbf{0.3683}                         \\ \cmidrule(l){2-6} 
                       & w/o Correction  & 0.5318                  & 0.3408                  & 0.6710                   & 0.3594                   \\
                       & w/o Item-side  & 0.5399                  & 0.3424                  & 0.6908                   & 0.3628                   \\
                       & w/o User-side  & 0.5377                  & 0.3419                  & 0.6870                   & 0.3620                   \\ 
                       & w/o Sampling     & 0.5420         & 0.3426         & 0.6898          & 0.3621          \\
                      \midrule
\multirow{6}{*}{NeuMF} & DCD     & \textbf{0.5287}         & \textbf{0.3303}         & \textbf{0.6734}          & \textbf{0.3497}                    \\ \cmidrule(l){2-6} 
                       & w/o Correction  & 0.5147                  & 0.3214                  & 0.6704                   & 0.3424                   \\
                       & w/o Item-side  & 0.5180                  & 0.3210                  & 0.6716                   & 0.3417                   \\ 
                       & w/o User-side  & 0.5225                  & 0.3254                  & 0.6681                   & 0.3450                   \\ 
                       & w/o Sampling     & 0.5196         & 0.3230         & 0.6706          & 0.3433          \\ \bottomrule
\end{tabular}
\label{table:ablation}
\vspace{-0.4cm}
\end{table}
\begin{figure}[h]

\centering
  \includegraphics[width=78mm]{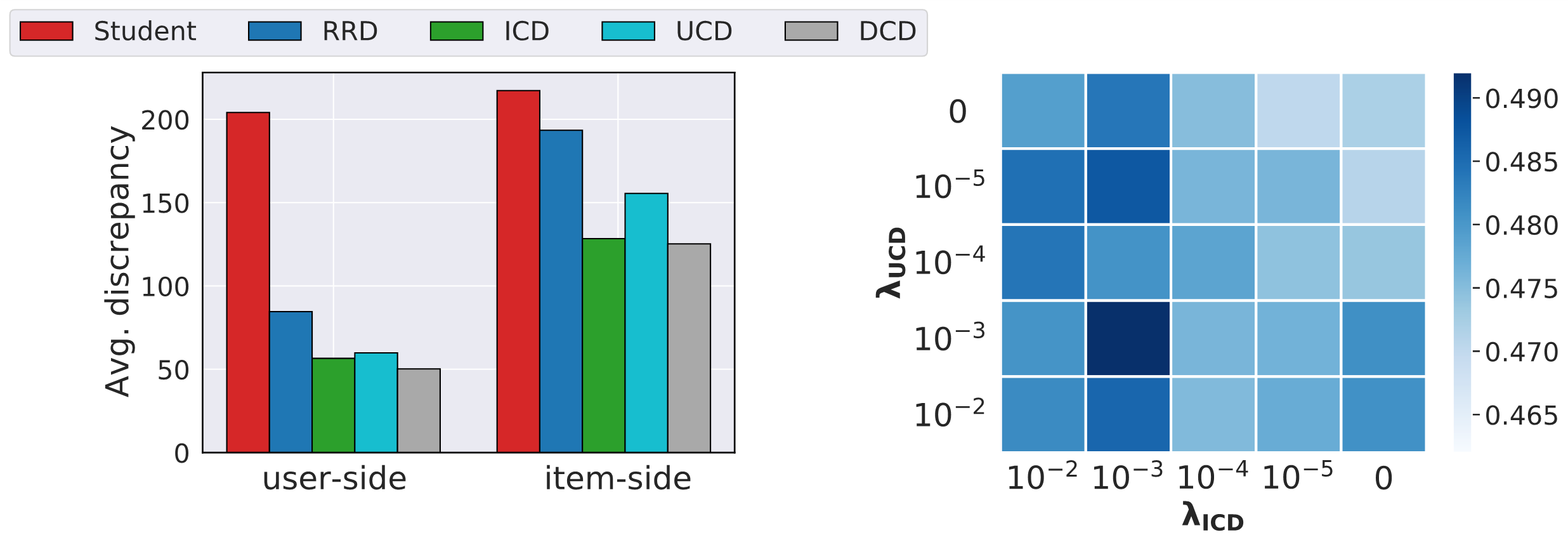}
  \vspace*{-5pt}
  \includegraphics[width=78mm]{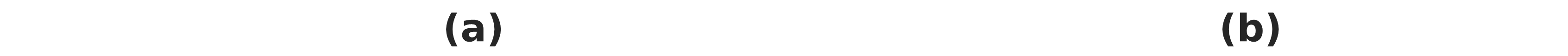}
  \vspace{-5pt}
  \caption{Effects of DCD.
  (a) The average discrepancy from Teacher,
  (b) H@5 with varying $\lambda_{UCD}$ and $\lambda_{ICD}$.}
  \label{fig:figure1}
 \vspace*{-0.6cm}
\end{figure}

\subsubsection{\textbf{Further Analysis}}
We provide further analysis on DCD. For the sake of the space, we report the results of BPR on CiteULike. 

First, Figure \ref{fig:figure1}a presents the average ranking-discrepancy of various methods.
In specific, we compute the discrepancy as $|R^*_S(\cdot) - R^*_T(\cdot)|$ for the user-side (and for the item-side) top-50 recommendation list produced by the teacher model.
We compute it for all users (and for all items), then report the average value.
We observe that the proposed correction strategy effectively reduces the discrepancy between the teacher model and the student model.
All the correction-based methods (i.e., ICD, UCD, and DCD) achieves lower discrepancy than RRD.
Also, DCD achieves the lowest discrepancy in both user-side and item-side, which supports its superior recommendation performance.
This also again shows the importance of the dual-side ranking correction.
Lastly, Figure \ref{fig:figure1}b shows the effects of $\lambda_{UCD}$ and $\lambda_{ICD}$.
Note that $\lambda_{UCD}=0 \, \&  \, \lambda_{UCD}=0$ corresponds to RRD.
We again observe that both user-side and item-side corrections are indeed effective.
The best performance is achieved when $\lambda_{ICD}$ is around $10^{-2}$-$10^{-3}$ and $\lambda_{UCD}$ is around $10^{-2}$-$10^{-3}$.

\section{Conclusion}
We propose DCD, a dual correction strategy for ranking distillation in top-$N$ RS.
Unlike the existing method based on unilateral distillation, DCD provides guidance designed to correct the errors that the student model has failed to learn. 
By considering the prediction errors of the student model, DCD helps to find an effective path for the student model’s training. 
DCD also considers the user-side ranking and item-side ranking simultaneously, providing a comprehensive view to understand both users and items.
We validate the effectiveness of DCD with extensive experiments on real-world datasets.
Also, we provide in-depth ablation study to ascertain the validity of each proposed component.
For future work, we will investigate the effects of DCD on various base models.

\vspace{0.1cm}
\noindent
\textbf{Acknowledgement.}
The authors thank SeongKu Kang for contributing to the implementation and improvement of DCD.
This work was supported by IITP grant funded by the MSIT: (No.2019-0-00075, Artificial Intelligence Graduate School Program(KAIST)) and the ETRI: (Contract No. 21ZS1100).

\bibliographystyle{ACM-Reference-Format}
\bibliography{acmart}


\begin{thebibliography}{30}


\ifx \showCODEN    \undefined \def \showCODEN     #1{\unskip}     \fi
\ifx \showDOI      \undefined \def \showDOI       #1{#1}\fi
\ifx \showISBNx    \undefined \def \showISBNx     #1{\unskip}     \fi
\ifx \showISBNxiii \undefined \def \showISBNxiii  #1{\unskip}     \fi
\ifx \showISSN     \undefined \def \showISSN      #1{\unskip}     \fi
\ifx \showLCCN     \undefined \def \showLCCN      #1{\unskip}     \fi
\ifx \shownote     \undefined \def \shownote      #1{#1}          \fi
\ifx \showarticletitle \undefined \def \showarticletitle #1{#1}   \fi
\ifx \showURL      \undefined \def \showURL       {\relax}        \fi
\providecommand\bibfield[2]{#2}
\providecommand\bibinfo[2]{#2}
\providecommand\natexlab[1]{#1}
\providecommand\showeprint[2][]{arXiv:#2}

\bibitem[\protect\citeauthoryear{Bachrach, Finkelstein, Gilad-Bachrach, Katzir,
  Koenigstein, Nice, and Paquet}{Bachrach et~al\mbox{.}}{2014}]%
        {tree_RS}
\bibfield{author}{\bibinfo{person}{Yoram Bachrach}, \bibinfo{person}{Yehuda
  Finkelstein}, \bibinfo{person}{Ran Gilad-Bachrach}, \bibinfo{person}{Liran
  Katzir}, \bibinfo{person}{Noam Koenigstein}, \bibinfo{person}{Nir Nice},
  {and} \bibinfo{person}{Ulrich Paquet}.} \bibinfo{year}{2014}\natexlab{}.
\newblock \showarticletitle{Speeding up the xbox recommender system using a
  euclidean transformation for inner-product spaces}. In
  \bibinfo{booktitle}{\emph{RecSys}}.
\newblock


\bibitem[\protect\citeauthoryear{Chen, Choi, Yu, Han, and Chandraker}{Chen
  et~al\mbox{.}}{2017}]%
        {chen2017learning}
\bibfield{author}{\bibinfo{person}{Guobin Chen}, \bibinfo{person}{Wongun Choi},
  \bibinfo{person}{Xiang Yu}, \bibinfo{person}{Tony Han}, {and}
  \bibinfo{person}{Manmohan Chandraker}.} \bibinfo{year}{2017}\natexlab{}.
\newblock \showarticletitle{Learning efficient object detection models with
  knowledge distillation}. In \bibinfo{booktitle}{\emph{NeurIPS}}.
\newblock


\bibitem[\protect\citeauthoryear{Cho, Kang, Hyun, and Yu}{Cho
  et~al\mbox{.}}{2021}]%
        {ProxySR}
\bibfield{author}{\bibinfo{person}{Junsu Cho}, \bibinfo{person}{SeongKu Kang},
  \bibinfo{person}{Dongmin Hyun}, {and} \bibinfo{person}{Hwanjo Yu}.}
  \bibinfo{year}{2021}\natexlab{}.
\newblock \showarticletitle{Unsupervised Proxy Selection for Session-based
  Recommender Systems}. In \bibinfo{booktitle}{\emph{SIGIR}}.
\newblock


\bibitem[\protect\citeauthoryear{Furlanello, Lipton, Tschannen, Itti, and
  Anandkumar}{Furlanello et~al\mbox{.}}{2018}]%
        {self_distill1}
\bibfield{author}{\bibinfo{person}{Tommaso Furlanello},
  \bibinfo{person}{Zachary~C Lipton}, \bibinfo{person}{Michael Tschannen},
  \bibinfo{person}{Laurent Itti}, {and} \bibinfo{person}{Anima Anandkumar}.}
  \bibinfo{year}{2018}\natexlab{}.
\newblock \showarticletitle{Born again neural networks}.
\newblock \bibinfo{journal}{\emph{arXiv preprint arXiv:1805.04770}}
  (\bibinfo{year}{2018}).
\newblock


\bibitem[\protect\citeauthoryear{He, Liao, Zhang, Nie, Hu, and Chua}{He
  et~al\mbox{.}}{2017}]%
        {NeuMF}
\bibfield{author}{\bibinfo{person}{Xiangnan He}, \bibinfo{person}{Lizi Liao},
  \bibinfo{person}{Hanwang Zhang}, \bibinfo{person}{Liqiang Nie},
  \bibinfo{person}{Xia Hu}, {and} \bibinfo{person}{Tat-Seng Chua}.}
  \bibinfo{year}{2017}\natexlab{}.
\newblock \showarticletitle{Neural collaborative filtering}. In
  \bibinfo{booktitle}{\emph{WWW}}.
\newblock


\bibitem[\protect\citeauthoryear{Hinton, Vinyals, and Dean}{Hinton
  et~al\mbox{.}}{2015}]%
        {KD}
\bibfield{author}{\bibinfo{person}{Geoffrey Hinton}, \bibinfo{person}{Oriol
  Vinyals}, {and} \bibinfo{person}{Jeffrey Dean}.}
  \bibinfo{year}{2015}\natexlab{}.
\newblock \showarticletitle{Distilling the knowledge in a neural network}.
\newblock \bibinfo{journal}{\emph{NIPS}} (\bibinfo{year}{2015}).
\newblock


\bibitem[\protect\citeauthoryear{Hu and Li}{Hu and Li}{2018}]%
        {multi-object}
\bibfield{author}{\bibinfo{person}{Jun Hu} {and} \bibinfo{person}{Ping Li}.}
  \bibinfo{year}{2018}\natexlab{}.
\newblock \showarticletitle{Collaborative multi-objective ranking}. In
  \bibinfo{booktitle}{\emph{CIKM}}.
\newblock


\bibitem[\protect\citeauthoryear{Hu, Koren, and Volinsky}{Hu
  et~al\mbox{.}}{2008}]%
        {hu2008collaborative}
\bibfield{author}{\bibinfo{person}{Yifan Hu}, \bibinfo{person}{Yehuda Koren},
  {and} \bibinfo{person}{Chris Volinsky}.} \bibinfo{year}{2008}\natexlab{}.
\newblock \showarticletitle{Collaborative filtering for implicit feedback
  datasets}. In \bibinfo{booktitle}{\emph{ICDM}}.
\newblock


\bibitem[\protect\citeauthoryear{Kang, Hwang, Kweon, and Yu}{Kang
  et~al\mbox{.}}{2020}]%
        {DERRD}
\bibfield{author}{\bibinfo{person}{SeongKu Kang}, \bibinfo{person}{Junyoung
  Hwang}, \bibinfo{person}{Wonbin Kweon}, {and} \bibinfo{person}{Hwanjo Yu}.}
  \bibinfo{year}{2020}\natexlab{}.
\newblock \showarticletitle{DE-RRD: A Knowledge Distillation Framework for
  Recommender System}. In \bibinfo{booktitle}{\emph{CIKM}}.
\newblock


\bibitem[\protect\citeauthoryear{Kang, Hwang, Kweon, and Yu}{Kang
  et~al\mbox{.}}{2021a}]%
        {IRRRD}
\bibfield{author}{\bibinfo{person}{SeongKu Kang}, \bibinfo{person}{Junyoung
  Hwang}, \bibinfo{person}{Wonbin Kweon}, {and} \bibinfo{person}{Hwanjo Yu}.}
  \bibinfo{year}{2021}\natexlab{a}.
\newblock \showarticletitle{Item-side Ranking Regularized Distillation for
  Recommender System}.
\newblock \bibinfo{journal}{\emph{Information Sciences}}
  (\bibinfo{year}{2021}).
\newblock
\showISSN{0020-0255}
\urldef\tempurl%
\url{https://doi.org/10.1016/j.ins.2021.08.060}
\showDOI{\tempurl}


\bibitem[\protect\citeauthoryear{Kang, Hwang, Kweon, and Yu}{Kang
  et~al\mbox{.}}{2021b}]%
        {HTD}
\bibfield{author}{\bibinfo{person}{SeongKu Kang}, \bibinfo{person}{Junyoung
  Hwang}, \bibinfo{person}{Wonbin Kweon}, {and} \bibinfo{person}{Hwanjo Yu}.}
  \bibinfo{year}{2021}\natexlab{b}.
\newblock \showarticletitle{Topology Distillation for Recommender System}. In
  \bibinfo{booktitle}{\emph{KDD}}.
\newblock


\bibitem[\protect\citeauthoryear{Kang, Hwang, Lee, and Yu}{Kang
  et~al\mbox{.}}{2019}]%
        {SSCDR}
\bibfield{author}{\bibinfo{person}{SeongKu Kang}, \bibinfo{person}{Junyoung
  Hwang}, \bibinfo{person}{Dongha Lee}, {and} \bibinfo{person}{Hwanjo Yu}.}
  \bibinfo{year}{2019}\natexlab{}.
\newblock \showarticletitle{Semi-supervised learning for cross-domain
  recommendation to cold-start users}. In \bibinfo{booktitle}{\emph{CIKM}}.
\newblock


\bibitem[\protect\citeauthoryear{Kang and McAuley}{Kang and McAuley}{2019}]%
        {candidategeneration}
\bibfield{author}{\bibinfo{person}{Wang-Cheng Kang} {and}
  \bibinfo{person}{Julian McAuley}.} \bibinfo{year}{2019}\natexlab{}.
\newblock \showarticletitle{Candidate Generation with Binary Codes for
  Large-Scale Top-N Recommendation}. In \bibinfo{booktitle}{\emph{CIKM}}.
\newblock


\bibitem[\protect\citeauthoryear{Kim, Lee, and Shim}{Kim et~al\mbox{.}}{2019}]%
        {dual_NPR}
\bibfield{author}{\bibinfo{person}{Seunghyeon Kim}, \bibinfo{person}{Jongwuk
  Lee}, {and} \bibinfo{person}{Hyunjung Shim}.}
  \bibinfo{year}{2019}\natexlab{}.
\newblock \showarticletitle{Dual neural personalized ranking}. In
  \bibinfo{booktitle}{\emph{WWW}}.
\newblock


\bibitem[\protect\citeauthoryear{Kingma and Ba}{Kingma and Ba}{2014}]%
        {kingma2014adam}
\bibfield{author}{\bibinfo{person}{Diederik~P Kingma} {and}
  \bibinfo{person}{Jimmy Ba}.} \bibinfo{year}{2014}\natexlab{}.
\newblock \showarticletitle{Adam: A method for stochastic optimization}.
\newblock \bibinfo{journal}{\emph{arXiv preprint arXiv:1412.6980}}
  (\bibinfo{year}{2014}).
\newblock


\bibitem[\protect\citeauthoryear{Kweon, Kang, Hwang, and Yu}{Kweon
  et~al\mbox{.}}{2020}]%
        {DRE}
\bibfield{author}{\bibinfo{person}{Wonbin Kweon}, \bibinfo{person}{Seongku
  Kang}, \bibinfo{person}{Junyoung Hwang}, {and} \bibinfo{person}{Hwanjo Yu}.}
  \bibinfo{year}{2020}\natexlab{}.
\newblock \showarticletitle{Deep Rating Elicitation for New Users in
  Collaborative Filtering}. In \bibinfo{booktitle}{\emph{WWW}}.
\newblock


\bibitem[\protect\citeauthoryear{Kweon, Kang, and Yu}{Kweon
  et~al\mbox{.}}{2021}]%
        {BD}
\bibfield{author}{\bibinfo{person}{Wonbin Kweon}, \bibinfo{person}{SeongKu
  Kang}, {and} \bibinfo{person}{Hwanjo Yu}.} \bibinfo{year}{2021}\natexlab{}.
\newblock \showarticletitle{Bidirectional Distillation for Top-K Recommender
  System}. In \bibinfo{booktitle}{\emph{WWW}}.
\newblock


\bibitem[\protect\citeauthoryear{Lee, Kang, Ju, Park, and Yu}{Lee
  et~al\mbox{.}}{2021}]%
        {BUIR}
\bibfield{author}{\bibinfo{person}{Dongha Lee}, \bibinfo{person}{SeongKu Kang},
  \bibinfo{person}{Hyunjun Ju}, \bibinfo{person}{Chanyoung Park}, {and}
  \bibinfo{person}{Hwanjo Yu}.} \bibinfo{year}{2021}\natexlab{}.
\newblock \showarticletitle{Bootstrapping User and Item Representations for
  One-Class Collaborative Filtering}.
\newblock \bibinfo{journal}{\emph{SIGIR}} (\bibinfo{year}{2021}).
\newblock


\bibitem[\protect\citeauthoryear{Lee, Choi, Lee, and Shim}{Lee
  et~al\mbox{.}}{2019}]%
        {CD}
\bibfield{author}{\bibinfo{person}{Jaewoong Lee}, \bibinfo{person}{Minjin
  Choi}, \bibinfo{person}{Jongwuk Lee}, {and} \bibinfo{person}{Hyunjung Shim}.}
  \bibinfo{year}{2019}\natexlab{}.
\newblock \showarticletitle{Collaborative Distillation for Top-N
  Recommendation}.
\newblock \bibinfo{journal}{\emph{ICDM}} (\bibinfo{year}{2019}).
\newblock


\bibitem[\protect\citeauthoryear{Lian, Liu, Ge, Zheng, Xie, and Cao}{Lian
  et~al\mbox{.}}{2017}]%
        {hash1}
\bibfield{author}{\bibinfo{person}{Defu Lian}, \bibinfo{person}{Rui Liu},
  \bibinfo{person}{Yong Ge}, \bibinfo{person}{Kai Zheng}, \bibinfo{person}{Xing
  Xie}, {and} \bibinfo{person}{Longbing Cao}.} \bibinfo{year}{2017}\natexlab{}.
\newblock \showarticletitle{Discrete Content-Aware Matrix Factorization}. In
  \bibinfo{booktitle}{\emph{KDD}}.
\newblock


\bibitem[\protect\citeauthoryear{Liu, He, Feng, Nie, Liu, and Zhang}{Liu
  et~al\mbox{.}}{2018}]%
        {hash2}
\bibfield{author}{\bibinfo{person}{Han Liu}, \bibinfo{person}{Xiangnan He},
  \bibinfo{person}{Fuli Feng}, \bibinfo{person}{Liqiang Nie},
  \bibinfo{person}{Rui Liu}, {and} \bibinfo{person}{Hanwang Zhang}.}
  \bibinfo{year}{2018}\natexlab{}.
\newblock \showarticletitle{Discrete factorization machines for fast
  feature-based recommendation}.
\newblock \bibinfo{journal}{\emph{arXiv preprint arXiv:1805.02232}}
  (\bibinfo{year}{2018}).
\newblock


\bibitem[\protect\citeauthoryear{Liu, Pham, Cong, and Yuan}{Liu
  et~al\mbox{.}}{2017}]%
        {Foursquare}
\bibfield{author}{\bibinfo{person}{Yiding Liu},
  \bibinfo{person}{Tuan-Anh~Nguyen Pham}, \bibinfo{person}{Gao Cong}, {and}
  \bibinfo{person}{Quan Yuan}.} \bibinfo{year}{2017}\natexlab{}.
\newblock \showarticletitle{An experimental evaluation of point-of-interest
  recommendation in location-based social networks}.
\newblock \bibinfo{journal}{\emph{Proceedings of the VLDB Endowment}}
  (\bibinfo{year}{2017}).
\newblock


\bibitem[\protect\citeauthoryear{Paszke, Gross, Massa, Lerer, Bradbury, Chanan,
  Killeen, Lin, Gimelshein, Antiga, et~al\mbox{.}}{Paszke
  et~al\mbox{.}}{2019}]%
        {paszke2019pytorch}
\bibfield{author}{\bibinfo{person}{Adam Paszke}, \bibinfo{person}{Sam Gross},
  \bibinfo{person}{Francisco Massa}, \bibinfo{person}{Adam Lerer},
  \bibinfo{person}{James Bradbury}, \bibinfo{person}{Gregory Chanan},
  \bibinfo{person}{Trevor Killeen}, \bibinfo{person}{Zeming Lin},
  \bibinfo{person}{Natalia Gimelshein}, \bibinfo{person}{Luca Antiga},
  {et~al\mbox{.}}} \bibinfo{year}{2019}\natexlab{}.
\newblock \showarticletitle{Pytorch: An imperative style, high-performance deep
  learning library}.
\newblock \bibinfo{journal}{\emph{arXiv preprint arXiv:1912.01703}}
  (\bibinfo{year}{2019}).
\newblock


\bibitem[\protect\citeauthoryear{Rendle, Freudenthaler, Gantner, and
  Schmidt-Thieme}{Rendle et~al\mbox{.}}{2009}]%
        {BPR}
\bibfield{author}{\bibinfo{person}{Steffen Rendle}, \bibinfo{person}{Christoph
  Freudenthaler}, \bibinfo{person}{Zeno Gantner}, {and} \bibinfo{person}{Lars
  Schmidt-Thieme}.} \bibinfo{year}{2009}\natexlab{}.
\newblock \showarticletitle{BPR: Bayesian personalized ranking from implicit
  feedback}. In \bibinfo{booktitle}{\emph{UAI}}.
\newblock


\bibitem[\protect\citeauthoryear{Romero, Ballas, Kahou, Chassang, Gatta, and
  Bengio}{Romero et~al\mbox{.}}{2014}]%
        {FitNet}
\bibfield{author}{\bibinfo{person}{Adriana Romero}, \bibinfo{person}{Nicolas
  Ballas}, \bibinfo{person}{Samira~Ebrahimi Kahou}, \bibinfo{person}{Antoine
  Chassang}, \bibinfo{person}{Carlo Gatta}, {and} \bibinfo{person}{Yoshua
  Bengio}.} \bibinfo{year}{2014}\natexlab{}.
\newblock \showarticletitle{Fitnets: Hints for thin deep nets}. In
  \bibinfo{booktitle}{\emph{arXiv}}.
\newblock


\bibitem[\protect\citeauthoryear{Tang and Wang}{Tang and Wang}{2018}]%
        {RD}
\bibfield{author}{\bibinfo{person}{Jiaxi Tang} {and} \bibinfo{person}{Ke
  Wang}.} \bibinfo{year}{2018}\natexlab{}.
\newblock \showarticletitle{Ranking distillation: Learning compact ranking
  models with high performance for recommender system}. In
  \bibinfo{booktitle}{\emph{KDD}}.
\newblock


\bibitem[\protect\citeauthoryear{Wang, Chen, and Li}{Wang
  et~al\mbox{.}}{2013}]%
        {CiteULike}
\bibfield{author}{\bibinfo{person}{Hao Wang}, \bibinfo{person}{Binyi Chen},
  {and} \bibinfo{person}{Wu-Jun Li}.} \bibinfo{year}{2013}\natexlab{}.
\newblock \showarticletitle{Collaborative topic regression with social
  regularization for tag recommendation}. In \bibinfo{booktitle}{\emph{IJCAI}}.
\newblock


\bibitem[\protect\citeauthoryear{Wang, Lian, and Ge}{Wang
  et~al\mbox{.}}{2019}]%
        {GCN_distill}
\bibfield{author}{\bibinfo{person}{Haoyu Wang}, \bibinfo{person}{Defu Lian},
  {and} \bibinfo{person}{Yong Ge}.} \bibinfo{year}{2019}\natexlab{}.
\newblock \showarticletitle{Binarized collaborative filtering with distilling
  graph convolutional networks}.
\newblock \bibinfo{journal}{\emph{IJCAI}} (\bibinfo{year}{2019}).
\newblock


\bibitem[\protect\citeauthoryear{Xia, Liu, Wang, Zhang, and Li}{Xia
  et~al\mbox{.}}{2008}]%
        {xia2008list-wise}
\bibfield{author}{\bibinfo{person}{Fen Xia}, \bibinfo{person}{Tie-Yan Liu},
  \bibinfo{person}{Jue Wang}, \bibinfo{person}{Wensheng Zhang}, {and}
  \bibinfo{person}{Hang Li}.} \bibinfo{year}{2008}\natexlab{}.
\newblock \showarticletitle{Listwise approach to learning to rank: theory and
  algorithm}. In \bibinfo{booktitle}{\emph{ICML}}.
\newblock


\bibitem[\protect\citeauthoryear{Zhang, Shen, Liu, He, Luan, and Chua}{Zhang
  et~al\mbox{.}}{2016}]%
        {DCF}
\bibfield{author}{\bibinfo{person}{Hanwang Zhang}, \bibinfo{person}{Fumin
  Shen}, \bibinfo{person}{Wei Liu}, \bibinfo{person}{Xiangnan He},
  \bibinfo{person}{Huanbo Luan}, {and} \bibinfo{person}{Tat-Seng Chua}.}
  \bibinfo{year}{2016}\natexlab{}.
\newblock \showarticletitle{Discrete Collaborative Filtering}. In
  \bibinfo{booktitle}{\emph{SIGIR}}.
\newblock


\end{thebibliography}

\end{document}